\documentclass[french, american, a4paper, 12pt]{article}%
\pdfoutput=1
\usepackage[T1]{fontenc} 
\usepackage{lmodern}

\usepackage{babel}
\usepackage{hyperref}
\usepackage{csquotes}
\usepackage{xspace}
\usepackage{microtype}
\usepackage[dvipsnames]{xcolor}
\usepackage{graphicx}

\usepackage[
	backend=biber,
	backref=false,
	sorting=nyt,
	language=auto,
	dashed=false,
	citestyle=authoryear, 
	bibstyle=authoryear, 
	]{biblatex}
 
\title{Assyrian Merchants meet Nuclear Physicists: History of the Early Contributions from Social Sciences to Computer Science. The Case of Automatic Pattern Detection in Graphs (1950s–1970s)}
 
\author{Sébastien Plutniak}
\date{\small Centre Émile Durkheim -- \textsc{umr} 5116, Université de Bordeaux, Bordeaux, France \\ 
   Laboratoire  \textsc{Traces} --  \textsc{umr} 5608, Université Toulouse Jean Jaurès, Toulouse, France\\
   \url{sebastien.plutniak@posteo.net}}

\begin{document}
\maketitle

\tableofcontents
\pagenumbering{gobble} 

\begin{abstract}
\noindent  
Community detection is a major issue in network analysis. This paper combines a socio-historical approach with an experimental reconstruction of programs to investigate the early automation of clique detection algorithms, which remains one of the unsolved NP-complete problems today. The research led by the archaeologist Jean-Claude Gardin from the 1950s on non-numerical information and graph analysis is retraced to demonstrate the early contributions of social sciences and humanities. The limited recognition and reception of Gardin's innovative computer application to the humanities are addressed through two factors, in addition to the effects of historiography and bibliographies on the recording, discoverability, and reuse of scientific productions: 1) funding policies, evidenced by the transfer of research effort on graph applications from temporary interdisciplinary spaces to disciplinary organizations related to the then-emerging field of computer science; and 2) the erratic careers of algorithms, in which efficiency, flaws, corrections, and authors’ status, were determining factors.
    
\medskip

\end{abstract}

Over the last 20 years, the application of methods based on graph theory has become widespread in many scientific fields, including the social sciences and humanities (hereafter SSH). The major proponents of these methods are physicists and computer scientists (e.g.  Albert-László Barabási, Theodore Lewis), who recognize a specific scientific domain they term  \enquote{network science}. 
 However, in contrast to contemporary narratives, the use of graphs (hereafter a synonym of networks) and more generally of algorithms are far from being new in  SSH. 
The current trend in network science is to dismiss the early applications and improvements of the graph analysis methods that were promoted by SSH researchers from the 1950s.
This paper addresses  how these early applications occurred and the reasons why they might have been forgotten or dismissed.
Clique detection, a method to detect connected subsets of nodes in a graph, is an example of such early algorithmic innovation. The collective research conducted on this subject by the French linguist, archaeologist and documentation specialist Jean-Claude Gardin (1925–2013) is explored. 
Recent research has started to emphasise the major role Gardin played in the second half of the  20\textsuperscript{th} century in the applications of information retrieval, automatic discourse analysis, and expert systems  (\cite{Dallas2015}) in SSH, and more generally in epistemology  (\cite{Plutniak2017jcpjcg-en}). 

\section{The clique problem}
\begin{figure}
	\begin{center}
		\includegraphics[width=.6\textwidth]{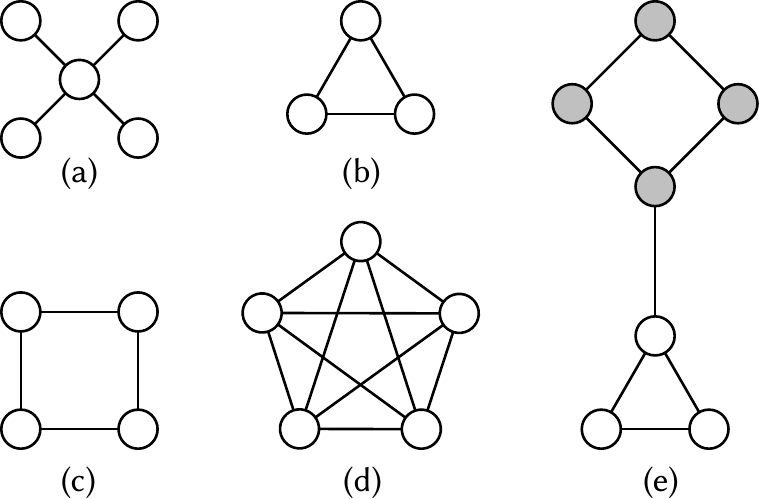}
		\caption{Examples of graph motifs discussed in the text:  star (a), cycles (b, c), cliques (b, d), a graph with two communities (e).}
		\label{fig:graph-motifs}
	\end{center}
\end{figure}

When a phenomenon is modelled by a graph (e.g. friendship relationships between individuals,  relationships between proteins, interactions among animals) detecting  particularly dense parts of the graph can be of interest (Figure~\ref{fig:graph-motifs}).
Many of the modern methods to achieve this are based on detecting classes of nodes, by determining a higher density of edges  (a class of methods nowadays commonly called \enquote{community detection}\footnote{Many of these methods rely on the optimization of modularity  (\cite{NewmanEtal2004}).}) or the identification of  elementary patterns, such as triadic configurations (\cite{WassermanFaust1994}).
A clique is another example of such specific patterns, which is of particular interest due to its early definition in the development of graph theory and applications.

The concept was first formally defined in a paper published by two mathematician psychologists in the \emph{Psychometrika}  journal (\cite{LucePerry1949}).
Later, the French psychologist and graph theorist specialist Claude Flament (1930--2019) reminded that a clique \enquote{is a systematization of a current notion in \emph{sociometry}: all the individuals of a clique choose each other.} (\cite[37]{Flament1965}).
Nevertheless, 
Luce, Perry, and Flament were all mathematicians committed to applying mathematics to psychological and sociological phenomena. The research fields of psychometry and sociometry occasionally overlapped and were particularly invested in by mathematicians (\cite{Freeman2004}).  
Sociometry refers to a field of inquiry originally coined in the early 1930s by  the Romanian-American  psychologist Jacob Levy Moreno (1889--1974).
It grew from his 1934 major publication \emph{Who shall Survive? A New Approach to the Problem of Human Interrelations} illustrating the use of graphs to represent and analyse empirically observed relationships  and  then from the \emph{Sociometry} journal (published 1937--1977).  

 From a mathematical perspective, the common definition of a clique is a \enquote{maximal complete subgraph of a graph} (\cite{MoonMoser1965}); or, formally:
\begin{quote}
$G' = (X'; V')$, a subgraph of $G = (X; V)$, is a clique if $(x, y) \in V' \Leftrightarrow x$  and $y \in X'$, that is, if all the possible arcs exist in $G'$. (\cite[37]{Flament1965})
\end{quote}
As a  clique can contain a clique, Flament also distinguished the concept of a maximal clique, \enquote{a clique which is not a subgraph of a clique}. 
Note that the publication of formal definitions did not limit variations in the use and semantics of the clique concept in the following decades; for example, defining a clique as a \enquote{subset of members who are more closely identified with one another than they are with the remaining members of their group} (\cite[377]{Hubbell1965}), i.e. as a synonym of the current concept of \enquote{community}\footnote{See also \cite{Lankford1974} for a review of detection methods where \enquote{clique} means \enquote{community}.}.

Despite the simplicity of the definition of cliques, automatizing their detection  through an algorithm raises serious computational difficulties, mainly due to  combinatorial explosion. Clique detection was included among the 21 NP-complete problems identified by Richard Karp (1935--), namely problems which are not computationally solvable in a deterministic time (\cite{Karp1972}). In this context, the research conducted from the late 1950s by Gardin and his collaborators was one of the early efforts to automatize clique detection.

\section{Gardin and Garelli's study of the Ancient Assyrian trade network}
In 1961, Gardin and the Assyriologist Paul Garelli (1924–2006) published probably the first automated application of graph theory to analyse historical materials, in the  \emph{Annales} journal (\cite{GardinGarelli1961}). They aimed to reconstruct a commercial network that was active in the 19\textsuperscript{th} century BC in present-day Cappadocia and, to do so, to:
1) automatize the deduction of the geographical locations of the merchants mentioned; 2) identify the general structure of the network of commercial relationships and the merchant groups; 3) discuss the interpretation of the synchronic and diachronic aspects of the resulting graph; 4) propose readable graphic representations of this graph; and 5) determine the degree and type of specialization of each merchant. The objectives, covering most of today’s major aspects of empirical graphs analysis, and the dataset used in this study, were unusual at the time, due to their ambition and size respectively.

 \begin{figure}
	\begin{center}
		\includegraphics[width=1\textwidth]{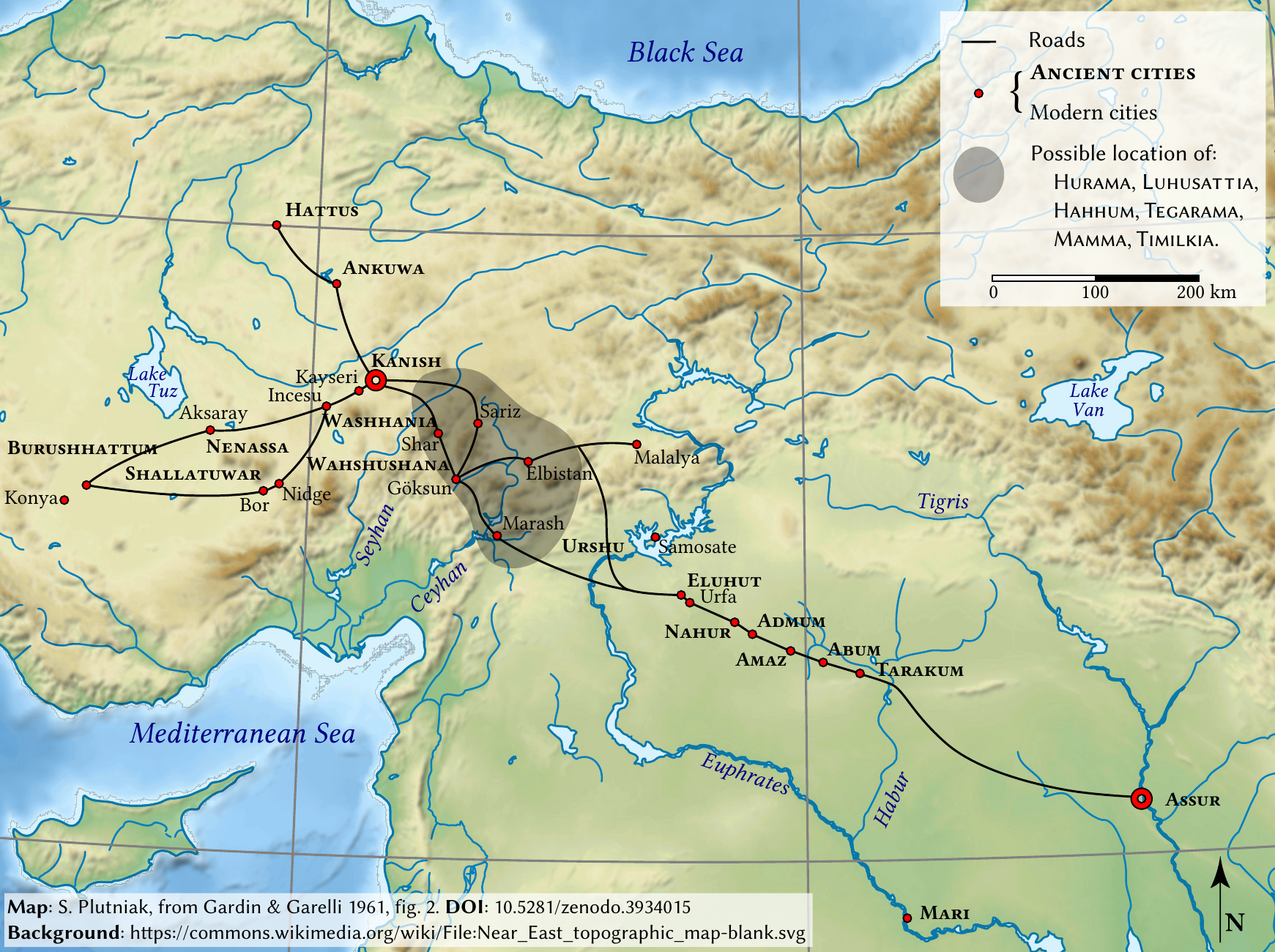}
		\caption{Assyrian commercial expansion during the 19\textsuperscript{th}~century BC (redrawn and adapted from Figure~2 from \protect\cite{GardinGarelli1961}).}
		\label{fig:map-assyrian-expansion}
	\end{center}
\end{figure}

The reconstruction of the Assyrian trade network was based on the corpus of cuneiform tablets found at the Kültepe site (near Kayseri, see Figure~\ref{fig:map-assyrian-expansion}).
These tablets have two characteristics: first, they document with text very ancient human activities (the 2\textsuperscript{nd} millennium BC), and second, there is a large amount of information available because these tablets have been collected, translated and published since the 19\textsuperscript{th} century, first by orientalists and then by specialized Assyriologists.
In the early 1920s, about 500 tablets had been published; by the end of the 1950s, this number had risen to about 2600.
This overabundance of material was an obstacle for research\footnote{For a detailed account of these data, see \cite[7--11]{Plutniak2018arcs}.}.

Gardin considered overcoming it by data sampling and the use of computer.
Sampling was carried out on two levels: first the tablets (the texts) and then the names of the identified merchants. At a meeting held in March 1961, Gardin indicated that the completion of the project presupposed the coding of the 2000 tablets studied by Garelli. The results which had been obtained so far were based on about 200 of them\footnote{\enquote{Comité de direction. Réunion du 16 mars 1961}, 16-03-1961, JCG~1, \textsc{Mae}.}.
Nothing was mentioned about the criteria used to select the tablets to be analysed. In a presentation in Paris at the   \enquote{\emph{Séminaire sur les modèles mathématiques dans les sciences sociales}}, Gardin announced  the number of \enquote{approximately} 1000 tablets (\cite[23]{Gardin1961b}).
The number of merchants’ names stated in the tablets was estimated at 1500 in the 1961 article, although the authors pointed out the possible problems of homonymy\footnote{\cite[847]{GardinGarelli1961}. However, in the  \enquote{\emph{Séminaire sur les modèles mathématiques dans les sciences sociales}}, Gardin mentioned \enquote{about} 3000 names  (\cite[23]{Gardin1961b})  and, in the later republication of the \emph{Annales} article in English, the number of 20,000 is given (\cite[380]{Gardin1965b}).}.  
A given merchant was defined by the number of transactions in which they were involved, since the authors assumed that it was a robust approximation of their relative importance  (\cite[854]{GardinGarelli1961}). 
In the 1961 article, the analysis focused on the thirty most important merchants\footnote{\cite[25]{Gardin1961b}.  The authors expected that subsequent analyses would have to cover about 100 names (\cite[876]{GardinGarelli1961},  \cite[457]{Gardin1962}). In a later commentary, Gardin mentions calculations made on a square matrix corresponding to the relationships among 200 merchants (\cite [389]{Gardin1965d}).} and the early clique detection algorithm was applied to this subset.

Paradoxically, despite the significance of the data analysed and its method, this study is rarely cited in the historical and archaeological literature on the ancient Orient, and never cited for its methodological dimension\footnote{For a detailed bibliographic study of its reception, see \cite[30--37]{Plutniak2018arcs}.}. It is also absent from the few publications available on the history of graph algorithms (\cite{Lankford1974}, \cite{Freeman1988}, \cite[201--205]{BarthelemyGuenoche1988},  \cite{PardalosXue1994}). Moreover, the available reports on this history are limited to the chronology of the different methods, without consideration of their social and intellectual contexts.  The rare citations of this research in subsequent studies contributed to its invisibility and other factors must be considered.

From the perspective of the social and intellectual history\footnote{Archives consulted: 
Jean-Claude Gardin's files at the Maison Archéologie-Ethnologie, Paris~X University,  Nanterre, France (hereafter abbreviated \textsc{Mae}); 
 \textsc{Euratom} files of the Historical archives of the European Commission, in Fiesole, Italy (\textsc{Haeu}) ;
\textsc{Cada} files at the \textsc{Cepam} archaeology laboratory, Nice, France (\textsc{Cepam});
files of the Institut Blaise Pascal,  private files of Pierre-Éric Mounier-Kuhn (\textsc{Ibp}).
} of computing (\cite{DemolPrimiero2014}), 
 the first part of this paper reports on the scientific significance of Gardin's research. Two factors which limited the subsequent reception and development of these applications in the SSH are then addressed:
1) an organizational factor, related to the lack of interest of historians and philologists in this approach and, consequently,  the transfer of Gardin's research program on graphs from a SSH research institution to applied mathematics and physics institutions;
2) an epistemic factor, related to the performances and potential error of the different algorithms, which also determine whether they are used by researchers.

\section{An organizational move from an interdisciplinary to a computing science space}

The analysis of the Assyrian trade network was the result of cooperation between two new research organizations focused on methods and interdisciplinarity, the \emph{Centre d’analyse documentaire pour l’archéologie} (\textsc{Cada}) led by Gardin in Paris, and the  European Scientific Information Processing Center (\textsc{Cetis}), a service created in 1961 as part of  the Joint Research Centre  of the European Atomic Energy Community (\textsc{Euratom}). After presenting these organizations and their collaboration, this section  shows 1) how the development of non-numerical data processing was the common background and motivation for this cooperation and 2) how this unusual interdisciplinary effort terminated, to be pursued by organizations dedicated to computer science.

\subsection{A precarious interdisciplinary methodological space}
 
\subsubsection{The \emph{Centre d'analyse documentaire pour l'archéologie}}
The \textsc{Cada}  was a laboratory of the French National Centre for Scientific Research  (\textsc{Cnrs})  created on the initiative of Gardin, located in Paris and Marseille. Despite its name, it was not restricted to archaeology but focused on methodological research  in information retrieval and mathematics applied to a wide range of disciplines,  including archaeology, history, literature, philology, linguistics, and medicine.

The composition of its scientific direction board reflects the uncommon nature of this research organization, in terms of its scientific themes and position in the institutional organization of scientific research in France\footnote{For a detailed description, see \cite[11--14]{Plutniak2018arcs}.}. It gathered leading figures in French   science policy, famous orientalists (e.g. Claude Schaeffer), open-minded methodological innovators (e.g. Henri Seyrig, Claude  Lévi-Strauss).
These actors were politically influential and shared Gardin’s ambitious methodological projects; they also explain the possibility and the funding of the \textsc{Cada}.
In addition, Gardin benefited from privileged relations with some members of the 5\textsuperscript{th} and 6\textsuperscript{th} sections of the \emph{École pratique des hautes études} (\textsc{Ephe}), in particular with some of the creators of the \emph{Maison des sciences de l'homme} (\textsc{Msh}), including Clemens Heller, 
 a close collaborator of Fernand Braudel. 
   The \textsc{Msh} was  then in its early stages of development, in the form of an association called  \emph{Association Marc Bloch},
what facilitated Gardin to obtain a research contract between this association and the \textsc{Euratom}  Joint Research Centre in 1959, allowing him to develop his work on computers.

\subsubsection{The research contract between the Marc Bloch association  and \textsc{Euratom}}
  
In July 1959, the \textsc{Euratom} created a  \emph{ Groupe de Recherche sur l'information scientifique automatique} (\textsc{Grisa}),  under the direction of the mathematician Paul Braffort (1923--2018).  
A policy of funding research by contract was immediately developed\footnote{As testified by the 22 contracts listed in \textsc{Meyer-Uhlenried}, Karl-Heinrich,  \enquote{organization de la collaboration entre la Commission et les institutions scientifiques des pays de la Communauté pour l'élaboration d'un langage documentaire (principe des contrats)}, 15 January 1960, BAC-059-1980 0209,  \textsc{Haeu}.}. 
Various scientific fields  were involved, including  \enquote{sociology}\footnote{Broadly understood to include physio-psychology.}: a project on the  \enquote{\emph{Traitement automatique de l'information dans les sciences humaines}}\footnote{Contract 001-60-1 CETF from the 10 March 1960.} was funded, under the direction of  Gardin.  
The  initial budget was 227,400~francs\footnote{\enquote{Contrat de recherche entre la Communauté européenne de l'énergie atomique et l'Association Marc Bloch},  21 December 1959, BAC-118-1986 1442, \textsc{Haeu}.}, 
(388,948~€ in 2019\footnote{Equivalent purchasing power in euros in 2017, taking into account monetary erosion due to inflation, calculated using the \textsc{Insee} converter (\url{https://insee.fr/fr/information/2417794}.}).
The purpose of the research contract was to develop an automatic documentation system based on the premise that 
\enquote{the language of the social sciences differs little from natural language, which is used in all exact sciences to communicate results}\footnote{Gardin J.‑C., \enquote{Programme d'études sémiologiques et documentaire (1961--1965)}, October 1960,  box Gardin~6, \textsc{Cepam}, p.~3.}, thus justifying the support of \textsc{Euratom}.
The research group led by Braffort aimed to develop a universal documentation language. Its main applications would be in the priority fields of \textsc{Euratom}, namely physics and mathematics, fields where scientific expression is highly standardized.  Braffort and Gardin believed that if it was possible to develop a documentation language for scientific fields where expression is much less standardized and more dependent on natural language (such as sociology), then this language would be \emph{a fortiori} applicable to the less complex  sciences whose expression is more standardized.

The calculations relating to the Assyrian tablets did not directly concern this project, which created the \textsc{Syntol} documentation language.
However, the calculations were possible due to the relational and financial resources obtained through this contract since the cost of the computer calculation hours had to be financed  (in this case, 21.52 machine hours on IBM 650 computers)\footnote{\enquote{Travaux du CETIS}, p.~8--9,   28 November 1960,  BAC-118-1986 1431, \textsc{Haeu}.}.
  
\subsubsection{Accessing cutting-edge computational resources}
Gardin’s approach consisted in systematically analysing the entities mentioned in the Akkadian texts, then coding them onto punched cards. There was a large number of descriptors, as in a similar repository previously planned and about which Gardin wrote:
\begin{quote}
[…] by the free combinatorial game of about five hundred terms, referring  to the fundamental elements of the natural environment and human actions --concrete (e.g. technology) or abstract (e.g. functions, institutions)-- the researcher can discover, in a vast and dispersed literature, all the relevant passages that shed light on the question he formulated: the transport of wood between the Syrian coast and  Mesopotamian countries, penalties incurred by a slave and a free man for the same crimes and, disturbances caused by the demobilization of troops, etc.  (\cite[14]{Gardin1960c}.)
\end{quote}
Concerning the relationships between the Assyrian merchants, the information in each \enquote{economic affairs} documented in the Akkadian texts was coded using thirteen variables coded into the 80 columns of the punch cards\footnote{See the description of the variables in \cite[848--850]{GardinGarelli1961}.}.
The commercial relationships were then represented as a matrix and the purpose of the analysis was to automate the detection of  particular sub-graphs, including cycles, stars, and cliques.
Despite the relatively few numbers of affairs and merchants, this type of  analysis quickly raised combinatorial explosion, requiring the use of a computer.
The calculations were carried out by André Debroux (1932--) and Peter Ihm (1926--2014) (\cite[875]{GardinGarelli1961}). Debroux was working for IBM Belgium, while collaborating with the \textsc{Grisa}, and Ihm, a German statistician, was a member of the \textsc{Grisa}.
In 1960, the \textsc{Grisa} was located in Brussels and did not yet have its own computers. Hours of calculations were rented or loaned by other organizations (\cite[56]{BraffortGazzano1961}), including the Free University of Brussels (\textsc{Ulb}), which had an IBM~650 computer \footnote{Acquired in 1957 (\cite [159]{HalleuxXhayet2007}).}. The computer used to analyse the  Assyrian trade relations was therefore either that of \textsc{Ulb} or IBM Belgium.

After the 1961 publication, Gardin envisaged continuing the experiment by integrating more merchants, but he pointed out the limits due to the insufficient computer power in Belgium:
\begin{quote}
The calculation of these groups, based on the matrix of relationships observed between a hundred individuals taken two by two, is, however, a complicated operation, which already exceeds the capabilities of an average computer such as the  IBM~650. A new program is about to be completed, for a more powerful machine (IBM~7090), with the collaboration of the European Scientific Information Processing Center.
 (\cite[457]{Gardin1962}.)
\end{quote}
The mathematical and computational problem was thus \enquote{transmitted} to the \textsc{Grisa} researchers.
In 1961, this research group had become a component of the \textsc{Cetis}, which was established on the  \textsc{Euratom} complex at  Ispra, near Varese in the north of Italy,  under the direction of Braffort.
Thus, the research contract between the Marc Bloch association and the \textsc{Euratom} facilitated Gardin's access  to the computers used by the researchers at  \textsc{Grisa} and  to mathematicians able to define and program the calculation methods.

\subsection{The rise of non-numerical data processing}
\subsubsection{Early developments  in Europe}
In addition to his activities in SSH, Gardin became involved in the then emerging scientific community at the intersection of mathematics, logic, linguistics, and documentation. Gardin was a graduate in linguistics and conducted his early research in automatic documentation. From the end of the 1950s, he participated in the first European research groups devoted to information retrieval and, more generally, to non-numerical information. The \enquote{Leibniz Seminars} brought together the most advanced scholars in this new field, with three main lines of research: automation of proof demonstrations, games simulation, and automation of documentation and translation.

In his 1961 paper on the Assyrian trade network, Gardin wrote that the processes executed by the computer were  calculations, in the broad sense given to this word by logicians, but different from those relating to numerical operations  (\cite[838]{GardinGarelli1961}). 
A few years later, Braffort gave a more detailed distinction between numerical and non-numerical computing problems in his book on cybernetics:
\begin{quote}
In numerical problems, the algorithm contains many formulas, arithmetic or logical and, in the machine, one can take advantage of the existence of registers linked to arithmetic and logical structures (accumulators, index registers).

[…]

In non-numerical problems, operations to be performed on the data use the information taken from the basic documentation, and express properties of the problem, but not those directly related to those of arithmetic and algebraic logic (e.g. commutativity, associativity).
 (\cite[135]{Braffort1968}.)
\end{quote}
Braffort, as the director of the \textsc{Cetis}, was one of the initiators of the Leibniz seminars.
In 1960, Gardin participated in the meeting entitled  \enquote{\emph{Enseignement préparatoire aux techniques de la documentation automatique}}, organised in Brussels by a linguist from the \textsc{Cetis}, André Leroy\footnote{This teaching took place from 15 to 22 February, 1960. Non-numerical information was one of the topic discussed, as illustrated by J.~Poyen's presentation \enquote{\emph{Quelques problèmes posés par le traitement de l'information non numérique}}.}. 
The next year, Gardin presented a conference entitled   \enquote{\emph{L'analyse sémantique dans les langues naturelles et formalisées}} at the \emph{Non-Numerical data-processing Symposium}, held at the IBM centre in Blaricum (Netherlands),  again organised by the \textsc{Cetis}\footnote{The symposium took place from 24 to 27 April, 1961, AAC~117, \textsc{Mae}.}.
In 1961, he was recruited as a \emph{directeur d'études} at the \textsc{Ephe} and entitled his teaching   \enquote{\emph{Automatique non-numérique}}, thus referring to:
\begin{quote}
{the other and more recent branch [which] covers a less well-defined field: \enquote{non-numerical data processing}, \enquote{information storage and retrieval}), automation of work on scientific texts (translations, documentation, etc.), so-called \enquote{heuristic} studies in the United States, etc.}\footnote{Gardin, J.‑C., \enquote{Propositions pour un enseignement sur l'automatique non-numérique (1962)}, 1961, box Gardin~6, \textsc{Cepam}.}.
\end{quote}

Through this work, Gardin became familiar with the researchers in this emerging field and with the scientific perspectives they promoted. Some of these researchers were among the first proponents of innovative work –Bourbaki in mathematics, Yehoshua Bar-Hillel and Noam Chomsky in language theory, and Claude Lévi-Strauss in anthropology– but also among the first and most informed critics of this work. 
For example, Marcel-Paul Schützenberger (1920--1996)  distanced himself from the Bourbaki group\footnote {During WWII, Schützenberger  participated in the French Resistance  as Gardin,  joining London in 1943.},  Braffort and Leroy opposed  the theory of the linguist  Lucien Tesnière (1893--1954) and that of Noam Chomsky (1928--),  and Bernard Jaulin (1934–2010) and Jacques Roubaud (1932--)  criticized the use of mathematics by Lévi-Strauss.

Graph analysis was considered as one of the problems  raised by non-numerical information. For example, Jacques Arsac (1929--2014), director of the  \emph{Institut de programmation} in Paris, reminded  that under the \enquote{non-numerical information} title, they \enquote{mainly taught graph theory} (\cite{Arsac1988}).
Therefore, for  the researchers committed  to this field, such as Braffort and Jaulin, the analysis of the Assyrian trade network, for which Gardin had requested their assistance, was a case study directly related to their research concerns about non-numerical information.
 
\subsubsection{The Assyrian trade network  from the mathematicians' perspective}
The \textsc{Grisa} conducted research on numerical (in particular, the calculation of nuclear reactors) and   non-numerical information (in particular information retrieval).
In the 1960 organization chart, Peter Ihm is associated with the working team on \enquote {applied mathematics}\footnote{\enquote{Rapport GRISA}, no.~2,  May 1960, CEAB12-640, \textsc{Haeu}.}.
 The team’s report specifies the work on the  \enquote{Problem of the Assyrian merchants (within the framework of the Gardin contract)}:
\begin{quote}
The results of this study will be used as pilot examples for the more general problem of determining the degree of dependence, the point of gravity, for example, in a set of arbitrary objects which are related by quantifiable relationships.

We are looking for an automatic information extraction algorithm based on the possible relationship between the elements of each pair of objects.\footnote{From the chapter on Ihm's activities in \enquote{Travaux du CETIS}, 28 November 1960, BAC-118-1986 1431, \textsc{Haeu}.}
\end{quote}
The problem raised by the case of the Assyrian commercial network is reformulated into a more general mathematical problem by Ihm.
Research on  graphs was an important axis of investigation at the \textsc{Grisa}/\textsc{Cetis}. In a 1961 document entitled \enquote{\emph{Progrès de l'automatique appliquée au domaine du traitement des informations}}, the aims of this research were detailed in the  section \enquote{\emph{L'étude des ensembles de graphes}}:
\begin{quote}
Diagrams, stemma, correlograms, flowcharts of arithmetic machines, analog diagrams, for example,… are all productions of these algebraic structures recently studied under the name of \enquote{graphs}. It was interesting to study these algebraic structures whose importance was growing steadily.
\begin{itemize}
	\item to represent and process them in the machine;
	\item to define the distance of two graphs, whose applications are obvious in the numerical domain (approximate calculation) as in the non-numerical domain (synonymy);
	\item finally to estimate the statistical properties of a population of graphs.\footnote{\enquote{Réunion du comité scientifique et technique du 14-3-1961. Progrès de l'automatique appliquée au domaine du traitement des informations}, 20 February 1961, BAC-118-1986 1431, \textsc{Haeu}.}
\end{itemize}
\end{quote}

In summary, when the analysis of the Assyrian commercial network was conducted, Gardin believed that it demonstrated the interest of automating the processing of non-numerical information in SSH. Graph analysis as such was a scientific objective, but only from the point of view of mathematicians. This division grew in the following years.

\subsection{Graph research projects move towards computer science organizations}

Projects following the 1961 Assyrian trade network analysis illustrate a progressive move of the research projects on graphs to research organizations dedicated to applied mathematics and computing.

\subsubsection{A complementary study: social networks in the New Hebrides}
The methods developed for the Assyrian trade network were immediately  reused in the \textsc{Cada} for another project on contemporary social networks in the New Hebrides (South Pacific Ocean, then a French condominium). 

\begin{figure}
	\begin{center}
		\includegraphics[width=1\textwidth]{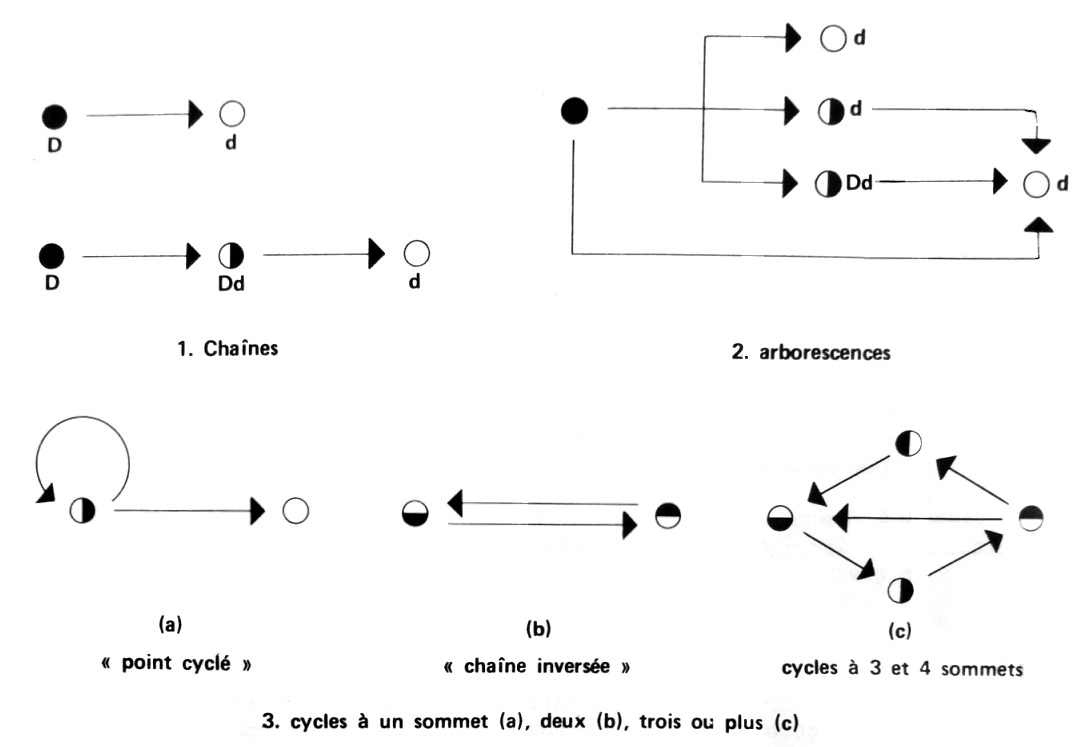}
		\caption{Patterns of interest in the study of the New Hebrides social network. (from \protect\cite[390]{EspiratEtal1973}). \enquote{D} stands for dominating titles and \enquote{d} for dominated titles.}
		\label{fig:espirat-etal1973p390}
	\end{center}
\end{figure}

The ethnographer Jean Guiart (1925–2019) conducted research on the local social structure, focusing on the system by which individuals could receive \enquote{titles} of prestige which were also related to places. Guiart collected a dataset of about 1200 titles and 500 places, related to individuals with about 2700 relations  (\cite[370]{EspiratEtal1973}).
 This research is first mentioned in relation to the \textsc{Cada} in 1962, after Guiart asked for support in running an experimental computer-based analysis of this large dataset (\cite{Gardin1962a})\footnote{\enquote{Orientation des travaux à partir du 2e semestre 1962}, June 1962, box Gardin~6, \textsc{Cepam}, p.~3.}. 
Two of Gardin's colleagues supervised this collaboration: Marie-Salomé Lagrange (1935--2011), a member of the \textsc{Cada}, and Monique Renaud, a member of another research group led by Gardin, the \emph{Section d'automatique documentaire}  (\textsc{Sad})\footnote{\cite[121]{NSF1959}.} at the \emph{Institut Blaise Pascal}, then the main place for computing research  in Paris.
A person was recruited to code the data on punched cards, Mrs A.-M.~Nougaret\footnote{As is often the case, very little information is available about subaltern research workers.}, preliminary tabulations were performed by  the \emph{Service mécanographique} of the \textsc{Cnrs}, and the main calculations were executed in 1963  on the IBM 1401 and 704 computers of the \emph{Institut Blaise Pascal} (\cite[387]{EspiratEtal1973}). Lagrange and Renaud wrote an internal report in 1965\footnote{Lagrange and Renaud. 1965. \emph{Étude d'un réseau sociologique aux Nouvelles-Hébrides, sur calculateur}, 55~p., mentioned in \cite{Gardin1965h}.} and the final results  were only published in 1973.
In addition, the analysis of the graphs includes detecting chains of different lengths, \enquote{arborescent} structures, and cycles of different lengths, but cliques were not studied  (Figure~\ref{fig:espirat-etal1973p390}).
As illustrated,  research organizations dedicated to computing started to play a major role, considering both the actors and the instruments.

 \subsubsection{The \textsc{Sad}'s programming language for graph analysis}
Based on the Assyrian and Polynesian projects, Gardin was keen to push forward their research on graphs. However, he favoured  the \textsc{Sad}, rather the \textsc{Cada} for this purpose, although he still led these two research organizations.
 In 1964, he obtained a grant from the \emph{Délégation générale à la recherche scientifique et technique} (\textsc{Dgrst}, a public scientific funding body) to conduct research on graphs at the   \textsc{Sad}\footnote{Contract  between the \textsc{Sad} and the   \textsc{Dgrst} n\textsuperscript{r}~64~FR~175. See also: \textsc{Sad}, \enquote{Rapport d'activités situation 1963--1964}, 6 March 1964, box Gardin~6, \textsc{Cepam}.}.
In a paper presenting the research conducted at the \textsc{Sad}, Gardin emphasized the reasons motivating the development of such a language:
\begin{quote}
However, programming the corresponding algorithms sometimes suffers from certain insufficiencies specific to the common symbolic languages, designed for numerical computation; then came the idea to define a programming language to specifically treat graph problems, our experience showed that they formed a large enough family to justify such a project. 
(\cite{Gardin1965h}.)
\end{quote}
Dirk Muysers, then an engineer-programmer at the \textsc{Sad}, was in charge of this research,  with the support of a young Greek mathematician recruited in 1965 for this purpose, Ion Arghiridis (1944--).
Muysers succeeded in creating the language, that he presented in a first report communicated to the \textsc{Dgrst} in 1966\footnote{Muysers D., \emph{Projet d'un langage de programmation numérique et non numérique, adapté plus particulièrement au traitement des données structurées par réseau}. Report  to the \textsc{Dgrst}, April 1966,  JCG~156, \textsc{Mae}.}. 
The report ended with two examples in which the language was used to implement a method to enumerate all the cycles in an oriented graph, and a method to find the shortest path between two nodes. However, this promising research quickly ended, for at least two identified reasons: Muysers stopped his research activities and left the \textsc{Cnrs} in 1968, and the \textsc{Dgrst} funded another project on graphs at the Paris \emph{Institut de programmation}.

\subsubsection{The \enquote{graph contract} of the \emph{Institut de programmation}}
The \emph{Institut de programmation} was founded in 1963 in  Paris and placed under the  direction of Jacques Arsac.
It was intended to be a counterpart to the \emph{Institut Blaise-Pascal} to develop the teaching of programming and computer science (\cite[428]{Mounierkuhn2012}).  
 However, research was also conducted at the \emph{Institut}, in particular on non-numerical information  and graphs. In 1964, the \emph{Institut} benefited from a  contract  on graphs funded by the \textsc{Dgrst}. This research  was carried out by  Jean Berstel (1941--) and  Jean-François Perrot  (1941--), two former Schützenberger's students, who co-signed the final report\footnote{Perrot J., Berstel, J.-F. 1967. \emph{Rapport final  de la  Convention  de  Recherche  DGRST  65 FR  002},  \enquote{Contrat Graphes}, mimeograph,  Institut  de  Programmation.}.

Arsac, Perrot and Berstel were trained in mathematics and dedicated to the different aspects of computer science. This contrasted with  \textsc{Cada}'s perspective where research in computing was a consequence of the data applications from the SSH.
Besides its strictly mathematical interest, research on graphs had practical consequences in the  domains where they can be applied, i.e. to optimize processes in operational research. 
Computer science was becoming a key domain for national development, as illustrated by the \emph{Plan Calcul} launched by De~Gaulle's government in 1966. 
This general context, and the reorientation of the \textsc{Dgrst} financial effort on  specialized research organizations, help to understand why Gardin and his collaborators abandoned their research on graphs. 
Subsequently, Gardin focused his research on information retrieval. He no longer referred to the analysis of the Assyrian trade network in his writings. He only reconsidered this study in the early 1970s when, turning his research towards the simulation of reasoning, he reinterpreted his early experiments on graphs as a prefiguration of his new interest in artificial intelligence\footnote{See \cite[26--30]{Plutniak2018arcs}.}.

\section{The efficiency, flaw, and career of algorithms}
After summarizing the chronology of clique detection methods, this section presents Bernard Jaulin and his algorithm. Then a report of our attempt in reconstructing and comparing this method to other algorithms is presented. The section ends with a discussion on flawed algorithms.

\subsection{A chronology of clique detection methods}
The concept of a clique had already been used in the sociometric literature   (e.g. \cite{ForsythKatz1946}) when it was first formally expressed  in 1949 by Robert Duncan Luce (1925–2012), an  American mathematician psychologist and games theorist   (\cite{LucePerry1949})\footnote{The same year a programmatic use of algebra to determine cliques was also published, although with no detailed procedure (\cite[156]{Festinger1949}).}.
From the early 1950s, procedures of matrix re-ordination were used to prove the structure of a graph, but not cliques in particular  (\cite{BeumBrundage1950}). 
The first algorithm was based on matrix algebra (\cite{HararyRoss1957}, hereafter Harary's method) and, two years later, a different method  based on the construction of   graphs in succession was proposed (\cite{PaullUnger1959}). 
However, these methods were theoretical and few or no implementations were published before the presentation of a  program for cluster detection\footnote{For a review, see  \cite{Lankford1974}.} (i.e. community detection), written for a \textsc{Univac}~1 machine by Duncan MacRae  and co-published with the sociologist James Coleman (1926--1995) (\cite{ColemanMacrae1960}).

This highlights the state-of-the-art when Gardin and Jaulin worked on the Assyrian trade network, for which the dataset size required a computer:
\begin{quote}
The mathematical problem of clique detection has been approached by many authors, from varying angles. The main issue, here, is in the size of the matrix which has to be analysed to that end (\emph{ca.} 200 rows x 200 columns); for any given detection method, it is unrealistic to consider its application on so large a matrix without the help of a computer. (\cite[389]{Gardin1965d})
\end{quote} 
Harary’s method was known by the authors, but they believed it could be improved in terms of efficiency and speed   (\cite[51]{Jaulin1961}, \cite[865]{GardinGarelli1961}).  So Jaulin developed his own algorithm.

\subsection{Bernard Jaulin and his  algorithm for clique detection}
In 1960, after becoming an engineer at the \emph{École des Arts et Métiers} and obtaining a bachelor of science degree, Jaulin joined the \enquote{\emph{Bureau d'études pour le traitement automatique de l'information dans les sciences humaines}}, created and led by  Gardin in the framework of the contract between the Marc Bloch association and the \textsc{Euratom}.
Gardin being a social scientist interested in formal approaches, and Jaulin being a mathematician interested in social sciences, they started a close collaboration also based on friendship.  
Together they organised in 1959 the first training program on non-numerical information processing at the   \emph{Institut Blaise Pascal}, and Gardin asked Jaulin to contribute to the study of the Assyrian trade network.
In 1964,  Jaulin became Director of the  Centre for Applied Mathematics of the \emph{Maison des sciences de l'homme} (former Marc Bloch association).
In July 1966, Jaulin and Gardin organised a conference entitled 
\enquote{\emph{Les applications du calcul dans les sciences de l'homme}}\footnote{Later published as \cite{GardinJaulin1968}.},   in Rome. This conference was supported by the UNESCO International Computing Centre (\textsc{Icc})\footnote{For a history of this centre, see \cite{Nofre2014}.}, also located in Rome and directed by the French mathematician Claude Berge (1926--2002). Berge was then a leading figure in graph theory and also  participated, as Gardin, in the  \enquote{Non-numerical data processing symposium} in Blaricum in 1961. 
Jaulin and Gardin's  conference in Rome was immediately followed by an  \enquote{International Seminar on Graph Theory and its Applications}, organised by Berge and the \textsc{Icc}, demonstrating the real interest for graph theory at the time.

  Jaulin presented his algorithm for clique detection in 1961 at the \enquote{\emph{Séminaire sur les modèles mathématiques en sciences humaines}} of the \textsc{Ephe} in Paris. 
The working   papers of this seminar are the only source  describing  this algorithm (\cite{Jaulin1961}). 
 A \textsc{Cada} internal report entitled \enquote{\emph{Sur une méthode de détermination des cliques dans un graphe symétrique}}  existed, but it has not been conserved in the archive repositories\footnote{It is indexed  but absent in the \textsc{Mae} archives, and not conserved   in the \textsc{Cepam} and \textsc{Haeu} archives}.
Jaulin's method was presented as an improvement of Harary's algorithm, using a different approach, namely boolean algebra, and included the notion of unicliqual points (reduction of the graph to treat nodes that belong to only one clique).
It was suitable for symmetric and unweighted graphs. 
Gardin pointed  out the last feature  as a possible flaw, because it did not consider different relationship intensities between merchants. 
However, he argued  that
1) they lacked a scale to decide the values used to weight the edges; and
2) they wanted to focus  on clique and cycle detection, without considering the weighting  of the graph  (\cite[864]{GardinGarelli1961}).
Jaulin’s algorithm was implemented twice, but neither programming code of implementation has been recovered. 
The first was in 1960, by  André Debroux (1932--?) from IBM Belgium and  Otto Hermann from the \textsc{Cetis}\footnote{\textsc{Cetis}, \emph{Travaux du CETIS}, 28 November 1960, BAC-118-1986 1431, \textsc{Haeu}.}. It was written in SOAP (Symbolic Optimal Assembly Program), the assembly language for the IBM~650 computer, and executed on this type of machine at IBM Belgium.
 The second was in 1962, for a IBM~704 computer\footnote{\textsc{Cnrs},  \enquote{Centre d'analyse documentaire pour l'archéologie} in \emph{Rapport d'activité. Octobre 1961--Octobre 1962}, p.~417.},  although it was probably abandoned.
 Consequently, we relied on the working paper to reconstruct Jaulin’s method.

 \subsection{Algorithm archaeology}
As archaeologists practising experimental archaeology to reproduce objects and their applications, computing historians have recently attempted to reconstitute computers and programs (\cite{DemolBullynck2008}, \cite{HaighPriestleyRope2014}).
Sometimes, computer scientists also perform these exercises. 
 For example, French mathematicians Jean-Pierre Barthélemy (1945--2010) and Alain Guénoche (1947--) implemented the Paull and Unger's algorithm for clique detection  thirty years after it was published in 1959  (\cite[201--205]{BarthelemyGuenoche1988}).
Note that from 1971, Guénoche was a member of the \textsc{Cada} and collaborated for about ten years in other research groups associated to Gardin.
 
\begin{figure}
	\begin{center}
		\includegraphics[width=.7\textwidth]{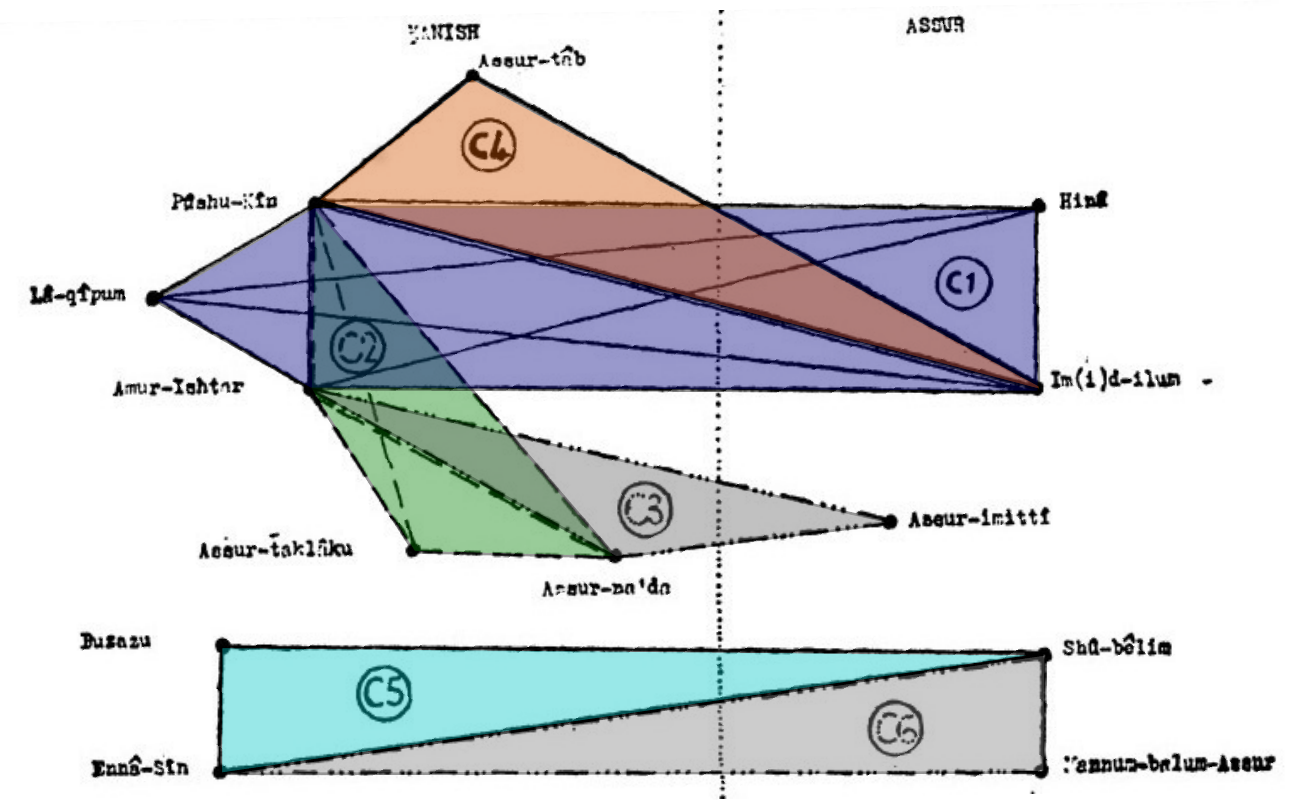}
		\caption{Illustration of  clique detection in the Assyrian trade network (Figure~10 in \protect\cite[868]{GardinGarelli1961}, adapted to highlight the cliques).} 
		\label{fig:GardinGarelli1961p868-color}
	\end{center}
\end{figure}

An attempt to implement Jaulin's  algorithm was carried out with the support of Guénoche. However, the description in the 1961 working paper was so incomplete and  muddled that it failed. 
A workaround was to extract the data sample presented in the paper on the Assyrian trade network to illustrate the detection of cliques and process it with other methods to compare the results. Some of these results are contradictory in this paper, e.g. the adjacency matrix presented in Figure~7 contains six cliques whereas the legend indicates three cliques and a singleton, which is actually part of a clique. 
Consequently, the data was extracted from a network plot 
 (Figure~\ref{fig:GardinGarelli1961p868-color}). 
Five methods were applied,
 including  Harary's algorithm, which was implemented in R language for this research\footnote{The algorithms used included:
 \cite{BronKerbosch1973}, implemented in the   \texttt{maxClique()} function from the \emph{RBGL} package;
\cite{Ostergard2001},  \texttt{qpGetCliques()}, \emph{qpgraph} package;  
 \cite{MakinoUno2004}: \texttt{clique.census()} function, \emph{sna} package; and
 \cite{EppsteinEtal2010},  \texttt{max\_cliques()} function, \emph{igraph} package. See supplementary materials.}.
All the methods identified the same six cliques as reported in Gardin and Garelli’s paper (Table~\ref{tab:comparaison}). 
However, Harary’s method surprisingly found two more cliques. This result and the contradictory legend of Gardin and Garelli’s paper drew our attention to errors in the use of formal methods.

\begin{table}
\caption{Identifier and size of the six cliques reported in \cite{GardinGarelli1961}, compared with the cliques detected by different methods.}\medskip
\begin{center}\small
	\begin{tabular}{llllllll}\hline
	 Id & Size & Gardin \& Garelli  & Harary  & Bron  & Makino  & Osertgard  & Eppstein   \\ 
	 &  (nodes)&  1961 &   1957 &  1973 &   2004 &  2001 &  2010 \\\hline
	1   & 5     & x                      & x           & x                     & x                  & x      & x             \\
	2   & 4     & x                      & x           & x                     & x                  & x         & x          \\
	3   & 3     & x                      & x           & x                     & x                  & x          & x         \\
	4   & 3     & x                      & x           & x                     & x                  & x         & x          \\
	5   & 3     & x                      & x           & x                     & x                  & x          & x         \\
	6   & 3     & x                      & x           & x                     & x                  & x           & x        \\
	–   & 3     & –                      & x           & –                     & –                  & –   & –             \\
	–   & 3     & –                      & x           & –                     & –                  & –     & –          \\\hline
	\end{tabular}\normalsize
\end{center}
\label{tab:comparaison}
\end{table}

\subsection{Flawed algorithms}
The different result generated by Harary's method led to suspect a flaw in the algorithm. Retracing the history of this method confirms  this hypothesis.
Frank Harary (1921--2005) was an American mathematician specialised in graph theory, with a great interest in developing applications for a wide range of domains, including social anthropology (\cite{HageHarary1983}).
After the publication of his method for clique detection in  1957 (\cite{HararyRoss1957}),  the method was extended to weighted graphs about 12 years later  (\cite{Doreian1969}). During that time, nobody saw that the 1957 algorithm was flawed before Harary himself noted it:
\begin{quote}
To set the record straight, that algorithm determines not only all the cliques of a graph, but sometimes a few other subgraphs as well. Correct algorithms for clique detection have subsequently been derived independently by several experts in computer programming. 
 (\cite[6]{Harary1970}.)
\end{quote}
Consequently, the method was then modified and  implemented in  the PL/1 language (\cite{Dixon1973}).
  This case illustrates how an algorithmic error can persist over a long time,  without limiting its  scientific spread (in January 2021, \emph{Google Scholars} reported 236 citations of the 1957 paper). 
 
 As we saw previously, clique detection was not applied in the study of the New Hebrides social network and was virtually forgotten by Gardin and his colleagues in their subsequent research. Similarly, Jaulin’s method is virtually absent from specialised literature on graphs, except a brief mention in Flament’s book on the \emph{Applications of Graph Theory to Group Structure} (\cite[37]{Flament1963})\footnote{Later literature reviews do not mention Jaulin's method, even those published in French, \emph{e.g.} \cite{Schneider1973}.}.
Contrary to the widespread use of Harary’s method, the flawed nature and the limited efficiency of Jaulin’s method might have finally prevented its author from publishing it. We have already mentioned how the only available description of the algorithm was muddled. In addition, Jaulin himself pointed out a limitation of his method, indicating that in this case Harary’s method must be used  (\cite[56]{Jaulin1961}).
The contrasting careers of these two flawed algorithms highlight the significant importance of external factors shaping the adoption and use of formal methods.
 
\section{Conclusion}
By combining a socio-historical approach with an experimental reconstruction this paper offers several contributions to the current development of computer science history. It brings to light forgotten early automated applications of graphs in the SSH developed in Europe, while emphasizing the role these disciplines played in this respect. The contribution of Gardin and his colleagues were particularly innovative. However, as illustrated by the case of the Assyrian trade network study,  their work received a very limited reception and little recognition in proportion to their originality and methodological value.
In addition to the effects of literature bibliographies on the discoverability and reuse of scientific productions, this paper discussed two possible factors for this under-valuation:
\begin{itemize}
	\item  the effects of research funding policies, demonstrated by the transfer of research effort on non-numerical information processing and graph applications from temporary interdisciplinary spaces (the \textsc{Cada} and the \textsc{Cetis}) to disciplinary organizations related to the then-emerging computer science;
	\item  the erratic careers of algorithms, in which efficiency, flaws, corrections, and authors’ status, were determining, but unpredictable factors.
\end{itemize}
These research case studies involved undeniably cutting-edge innovations for their time. Detailed analyses show that their long-term scientific value and effects are less controllable than what is sometimes assumed by science policy-makers who attempt to organise science as technological innovations are managed.

\section*{Acknowledgements}
I thank Pierre-Éric Mounier Kuhn for his advice and providing me with the archives of the \emph{Institut Blaise Pascal}, Solène Chevalier for her help in accessing some documents and Alain Guénoche for his support in understanding Jaulin’s algorithm.



\section*{Supplementary material}

\begin{itemize}
	\item \url{https://doi.org/10.5281/zenodo.3932104}
	\item \url{https://doi.org/10.5281/zenodo.3934015}
\end{itemize}

\printbibliography[title=References]

\end{document}